\documentstyle[12pt]{article}

\newcommand{\undersmash}[1]{
  \ifmmode \underline{\smash{\hbox{#1}}}
  \else $\underline{\smash{\hbox{#1}}}$
  \fi
}
\newenvironment{Definition}{\medskip\par\noindent {\bf Definition:}
\rm}{\medskip\par}
\newcommand{\REELL}{{\setlength{\unitlength}{1em}
                     \begin{picture}(0.75,1)
                     \put(0,0){\line(0,1){0.69}}
                     \put(0,0){\sf R}
                     \end{picture}
                   }}
\newcommand{\KOMPLEX}{{\setlength{\unitlength}{1em}
                     \begin{picture}(0.7,1)
                     \put(0.34,0){\line(0,1){0.65}}
                     \put(0,0){\sf C}
                     \end{picture}
                   }}
\newcommand{\NATUR}{{\setlength{\unitlength}{1em}
                     \begin{picture}(0.75,1)
                     \put(0,0){\line(0,1){0.69}}
                     \put(0,0){\sf N}
                     \end{picture}
                   }}
\newcommand{\FNATUR}{\mbox{\tiny{\setlength{\unitlength}{1em}
                     \begin{picture}(0.6,0.5)
                     \put(0,0){\line(0,1){0.48}}
                     \put(0,0){\rm N}
                     \end{picture}
                   }}}
\newcommand{\FKOMPLEX}{\mbox{\tiny{\setlength{\unitlength}{1em}
                               \begin{picture}(0.6,0.5)
                               \put(0.34,0){\line(0,1){0.47}}
                               \put(0,0){\rm C}
                               \end{picture}
                              }}}
\newcommand{\FREELL}{\mbox{\tiny{\setlength{\unitlength}{1em}
                     \begin{picture}(0.6,0.5)
                     \put(0,0){\line(0,1){0.48}}
                     \put(0,0){\rm R}
                     \end{picture}
                   }}}
\newcommand{\Brr}{{\mathchoice{\REELL}{\REELL}{\!\FREELL}{\!\FREELL}}}
\newcommand{\Bcc}{{\mathchoice{\KOMPLEX}{\KOMPLEX}{\!\FKOMPLEX}{
\!\FKOMPLEX}}}
\newcommand{\Bnn}{{\mathchoice{\NATUR}{\NATUR}{\FNATUR}{\FNATUR}}}
\newcommand{\separate}{\medskip\noindent}
\newcommand{\Lieb}{\bfb}
\newcommand{\Lieg}{\bfg}
\newcommand{\Liek}{\bfk}
\newcommand{\Liem}{\bfm}
\newcommand{\Lien}{\bfn}

\newcommand{\Lieu}{\bfu}
\newcommand{\Liesl}{\mbox{\bf sl}}

\newcommand{\Lieh}{{\mathchoice{\bfh}{\bfh}{\!\mbox{\tiny\bfh}}{
  \!\mbox{\tiny\bfh}}}}
\newcommand{\Liehatg}{{\hat\bfg}}
\newcommand{\Liechg}{{\check\bfg}}
\newcommand{\Lietilg}{{\widetilde\bfg}}

\newcommand{\LieG}{\bfG}
\newcommand{\LieH}{\bfH}
\newcommand{\LieK}{\bfK}

\newcommand{\LieN}{\bfN}
\newcommand{\LieGL}{\mbox{\sf GL}}
\newcommand{\LieSL}{\mbox{\sf SL}}

\newcommand{\LietilG}{\widetilde{\bfG}}
\newcommand{\LiechG}{\check{\bfG}}
\newcommand{\LiehC}{{\Lieh_\Bcc}}
\newcommand{\LiekC}{{\Liek_\Bcc}}
\newcommand{\LieKC}{{\LieK_\Bcc}}
\newcommand{\halb}{\frac{1}{2}}
\newcommand{\inv}{^{-1}{}}
\newcommand{\nC}{(n,\Bcc)}
\newcommand{\lpair}{\langle}
\newcommand{\rpair}{\rangle}

\newcommand{\bfk}{{\bf k}}

\newcommand{\bfb}{{\bf b}}

\newcommand{\bfg}{{\bf g}}
\newcommand{\bfh}{{\bf h}}
\newcommand{\bfm}{{\bf m}}
\newcommand{\bfn}{{\bf n}}

\newcommand{\bfu}{{\bf u}}

\newcommand{\bfG}{{\bf G}}
\newcommand{\bfH}{{\bf H}}
\newcommand{\bfK}{{\bf K}}

\newcommand{\bfN}{{\bf N}}

\newcommand{\calC}{{\cal C}}
\newcommand{\calD}{{\cal D}}

\newcommand{\calK}{{\cal K}}
\newcommand{\calL}{{\cal L}}

\newcommand{\calU}{{\cal U}}

\newcommand{\tilc}{{\tilde{c}}}

\newcommand{\tilg}{{\tilde{g}}}

\newcommand{\tilp}{{\tilde{p}}}
\newcommand{\tilq}{{\tilde{q}}}

\newcommand{\tilz}{{\tilde{z}}}

\newcommand{\tilC}{{\widetilde{C}}}

\newcommand{\tilG}{{\widetilde{G}}}

\newcommand{\hatg}{\hat{g}}

\newcommand{\hatx}{\hat{x}}

\newcommand{\chO}{\check{O}}

\newcommand{\unp}{\undersmash{p}}
\newcommand{\unt}{\undersmash{t}}
\newcommand{\diag}{\mbox{\rm diag}}

\newcommand{\bref}[1]{(\ref{#1})}

\newcommand{\Ad}{{\rm Ad}}
\newcommand{\TAd}{\widetilde{\Ad}}
\newcommand{\diff}{{\rm d}}
\newcommand{\DEF}{\stackrel{\rm def}{=}}
\newcommand{\Res}[2]{\mbox{\rm Res}_{#1}\left(#2\right)}
\newcommand{\height}{{\rm ht}}
\newcommand{\BEQ}{\begin{equation}}
\newcommand{\EEQ}{\end{equation}}
\newcommand{\BEA}{\begin{eqnarray}}
\newcommand{\EEA}{\end{eqnarray}}
\newcommand{\der}[2]{\frac{\diff{#1}}{\diff{#2}}}
\newcommand{\Drinfeld}{\mbox{Drinfel{\kern-0.3mm'\kern-0.5mm}d}}
\newcommand{\param}{s}
\newcommand{\Pq}{{\Phi_{q_0}}}
\newcommand{\Oq}{{\Omega_{q_0}}}

\newcommand{\DG}{\calD\LieG}
\newcommand{\Dg}{\calD\Lieg}
\newcommand{\LG}{\calL\LieG}
\newcommand{\BF}{Birkhoff factorization}
\newcommand{\Op}{{\Omega_{p_0}}}

\begin{document}
\title{Group Theoretical Symmetries and Generalized B\"acklund
Transformations for Integrable Systems}
\author{Guido Haak\\Institut f\"ur Theoretische Physik, FU-Berlin\\
Arnimallee 13, D-1000 Berlin 33}
\date{}
\maketitle

\begin{abstract}
We present a notion of symmetry for 1+1-dimensional integrable
systems which is consistent with their group theoretic description and
reproduces in special cases the known B\"acklund transformation for the
generalized Korteweg-deVries hierarchies.
We also apply it to the relativistic invariance of the Leznov-Saveliev
systems.
\end{abstract}
%
In \cite{HSS92} a general and detailed group theoretic description of
integrable systems, based on the work of Segal, Wilson, Reyman and
Semenov-Tian-Shansky was given.
It included the work of \Drinfeld\ and Sokolov on
the generalized modified Korteweg-deVries (mKdV) hierarchy and the notion
of a generalized Miura transformation for these systems.
The aim of this work is to interpret some of the well known properties of
integrable systems in this group theoretic setting.

As the relation between the
B\"acklund transformation, Miura transformations and symmetries of
differential equations was one of the hallmarks of the development of the
inverse scattering method~(for an account on this see e.g.~\cite{Newell}),
it is of interest to exhibit
the relationship between these structures and the group
theoretic approach. Thus an aim will be to
establish an appropriate notion of a symmetry and in particular to
explain the B\"acklund transformation of
the KdV models for the group $\LieSL\nC$ in
terms of the group theoretic setting.

The first section gives a short description of the group theoretic
formulation as presented in~\cite{HSS92}. The examples of the
generalized nonlinear Schr\"odinger (NLS) equation and the sine-Gordon
equations are then presented in more detail.

In the second section we present
the formulation of a symmetry and relate it by the
generalized Miura transformation of \Drinfeld, Shabat and Sokolov~\cite{SS,DS}
to the B\"acklund transformation in close analogy to the well known case of the
KdV hierarchy. We give
an explicit reduction for the group theoretic hierarchy of the
generalized NLS equation, which produces the mKdV flows, and
a B\"acklund transformation for the generalizations of the KdV equation
belonging to the groups $\LieSL\nC$.

At the end we show, how this kind of symmetry can be generalized to
include the Lorentz group symmetry of the Leznov-Saveliev systems.

\section{The Group Theoretical Setting}
Following \cite{HSS92} an integrable system is given in terms of
the following abstract data:
\begin{itemize}
\item[a)] An admissible triple of Lie-Groups
$(\LieG,\LieG_-,\LieG_+)$ with $\LieG_\pm\subset\LieG$ closed, the
multiplication map $\LieG_-\times\LieG_+\rightarrow\calU$
being a diffeomorphism into an open dense subset $\calU$ of $\LieG$.
This is an abstract Birkhoff factorization. In addition the
following requirements are made:
\begin{itemize}
\item[1)] The corresponding Lie algebras form a Manin triple
$(\Lieg,\Lieg_-,\Lieg_+)$, i.e.\ there exists a Killing form $B_G$ on
$\Lieg=\Lieg_-\oplus\Lieg_+$ such that $\Lieg_-$ and $\Lieg_+$ are
maximally isotropic w.r.t.~$B_G$.
\item[2)] The exponential map $\exp:\Lieg_-\rightarrow\LieG_-$ is bijective.
\end{itemize}
\item[b)] An element $p\in\Lieg_+$ is fixed.
\end{itemize}
The Birkhoff factorization~\cite{PS} is clearly modeled to generalize the
decomposition of $\LieGL\nC$ into upper and lower triangular matrices
to infinite dimensional groups. Note also that $\Lieg_+$
(and $\Lieg_-$) bears the additional structure of a Lie
Bialgebra~\cite{Drin} induced by the Manin triple $(\Lieg,\Lieg_-,\Lieg_+)$.

$p$, called the generalized momentum,
will play the r\^ole of the generator of space translations.
The $x$-space flow $g_-(x)$ in $\LieG_-$ for small $x\in\Brr$
through an element $g_-=g_-(0)\in\LieG_-$
is defined in the group theoretically most natural way via
\BEQ
g(x)=\exp(-x p)g_-=g_-(x)g_+(x),
\EEQ
where the second equation denotes the Birkhoff factorization of $g(x)$.

In the following we choose a Cartan subalgebra $\Lieh$ of $\Lieg$, such that
$p$
is an element of $\Lieh_+=\Lieh\cap\Lieg_+$.
We may then take any element of $\Lieh_+$ to generate further
local flows in the group $\LieG$, which commute with the $x$-flow.
We write them in the compact form
\BEQ \label{flows}
g(\unt)=\exp(-\lpair\unt,\unp\rpair)g_-,
\EEQ
where $\unt=(t_0=x,t_1,\ldots)$, $\unp=(p_0=p,p_1,\ldots)$. It is always
understood that the parameters $t_j\in\Brr$
are small, and only finitely many are nonzero.

In the forthcoming examples $\LieG$ is a loop group with infinite rank.
Therefore the systems described above possess infinitely many commuting
flows, giving also an infinite set of integrals of motion.

The connection with the realization of integrable systems as zero
curvature conditions (Zakharov-Shabat (ZS) type equations) is now given by
choosing the natural flat connection on the trivial principal bundle
$\LieG/\LieG_+\supset\LieG_-$, the pullback of the (right invariant)
Maurer-Cartan form on $\LieG_+$ with the map $g_+:\LieG\rightarrow\LieG_+$:
\BEQ
\omega=-\diff g_+g_+\inv.
\EEQ
The Maurer-Cartan equation,
\BEQ
\diff\omega+[\omega,\omega]=0,
\EEQ
translates into a zero curvature condition of the ZS type for the
components $V_j(\unt)$ of the connection in the direction of the
right invariant vector fields $r(p_j)$ for $p_j$:
\BEQ \label{UDEF}
V_j(\unt)=\omega(r(p_j))(g(\unt))=q_+(\Ad(g_-\inv(\unt))p_j),
\EEQ
\BEQ \label{ZS}
\der{V_k(\unt)}{t_j}-\der{V_j(\unt)}{t_k}-[V_k(\unt),V_j(\unt)]=0,
\EEQ
together with an auxiliary equation
\BEQ \label{aux}
g_-\inv(\unt)\der{g_-(\unt)}{t_j}=-q_-(\Ad(g_-\inv(\unt))p_j).
\EEQ
Here we denote by $q_\pm$ the projections $\Lieg\rightarrow\Lieg_\pm$.
$U(\unt)=V_0(\unt)$ is by~\bref{UDEF} and the fact that $\Lieg_-$ is
maximally isotropic w.r.t.~a nondegenerate form $B_G$, a generic element
of the orbit of the coadjoint action of $G_-$ on $\Lieg_+$ through $p$. The
Lie-Kirillov-Kostant-Souriau symplectic structure on this orbit gives a
Poisson structure being compatible with the one given by the Lie Bialgebra
$\Lieg_+$.

The equation~\bref{UDEF} gives a map
$\Op$ from the coset $\LieH_-(p_0)\setminus\LieG_-$, $\LieH_-(p_0)$ being
the centralizer of $p_0$ in $\LieG_-$, to the set $\calK_0(\Brr,\Lieg_+)$
of smooth functions defined in a neighborhood of the origin in $\Brr$
with values in $\Lieg_+$.
We assume this map to be injective. For a fixed admissible triple
$(\LieG,\LieG_-,\LieG_+)$ this is a restriction on the momentum operator $p_0$.

Restricting attention to the case of solutions $U(\unt)$ periodic in $x$
(say with period $l$), we may use the assumed injectivity of $\Op$ to
define for every periodic solution $U(x)$ a periodic element
$\tilg_-(x)$ of $\LieG_-$ by
\BEQ
\tilg_-(x)=\exp(-xI)g_-(x),
\EEQ
where $g_-(x)$ is any $x$-space flow in $\LieG_-$ mapped by $\Op$ to
$U(x)$ and $I\in\Lieh_-$ is defined by
\BEQ
g_-(l)=\exp(lI)g_-(0).
\EEQ
Notice that $g_-(x)$ and $\tilg_-(x)$ are in the same equivalence class
modulo $\LieH_-(p_0)$.

Combining equations~\bref{UDEF} and~\bref{aux} for $j=0$ this gives
\BEA
\omega(r(p_0))(g(x)) & = & \Ad(g_-\inv(x))p_0+
g_-\inv(x)\der{g_-(x)}{x} \nonumber\\
& = & \Ad(\tilg_-\inv(x))(p_0+I)+\tilg_-\inv(x)\der{\tilg_-(x)}{x}.
\EEA
Using Frenkel's definition~\cite{Frenkel} of the affine Lie group
$\LG$ connected with $\LieG$ this may be rewritten as
\BEQ
\omega(r(p_0))(g(\cdot))+1\calD=\TAd(\tilg_-(\cdot))(p_0+I+I_0\calC+1\calD)
\EEQ
where $I_0=-lB_G(p_0,I)$.

The periodic $x$-space flows may thus be incorporated into the group
theoretic scheme by substituting the admissible triple
$(\LieG,\LieG_-,\LieG_+)$ by a triple $(\LG,\LG_-,\LG_+)$ which is
also admissible~\cite[Lemma~2.6]{HSS92}.

As an example let us take the triple of analytic loop groups
\BEQ
(\LieG,\LieG_-,\LieG_+)=
(L^a\LieKC,L^{a-}_1\LieKC,L^{a+}\LieKC),
\EEQ
$\LieK$ a compact semisimple Lie group, $\LieKC$ its complexification
with Lie algebra $\LiekC$.

In order to achieve bijectivity of the exponential map
$\exp:\Lieg_-\rightarrow\LieG_-$, we don't use the usual construction of
a Loop group over $\LieKC$ but the one given in~\cite{HSS92}. There
the group $L^a\LieKC$ is defined to be the stalk at infinity
of the sheaf of analytic functions on $\Bcc$, i.e.\ $L^a\LieKC$
consists of functions which are defined and analytic outside some compact
set. The subgroup $L^{a+}\LieKC$ consists of the entire maps from
$\Bcc$ to $\LieKC$, and $L^{a-}_1\LieKC$ consists of
the maps, which are defined and analytic
in some open neighborhood of $\infty$ and have the value $1$ there.
Their Lie algebras are $(L^a\LiekC,L^{a-}_0\LiekC,L^{a+}\LiekC)$ with
self-explanatory notation.
The choice of the Killing form
\BEQ
B_{L^a\LieKC}(a(\cdot),b(\cdot))=\Res{\lambda=\infty}{B_\LieKC(a(\lambda),
b(\lambda))}
\EEQ
makes this triple of Lie algebras into a Manin triple.

For $p_0=\lambda A$, $A\in\LiehC$ with $\LiehC$ being
the Cartan subalgebra of $\LiekC$,
the resulting integrable system is the NLS hierarchy, generalizing the
vector-NLS of Fordy and Kulish~\cite{FK}.
By a reduction described in the next section we get the mKdV type systems.

For all these systems and all generic $p_0$, injectivity of $\Op$
holds and the
integrals of motion can be calculated explicitly, as can be easily seen
by generalizing the proof in \cite{HSS92} for $p_0=\lambda A$ to
arbitrary $p_0$.

A second important example is the sine-Gordon equation and its
generalizations, the Leznov-Saveliev systems~\cite{Leznov},
in light-cone variables.

They belong to the following construction, also considered in~\cite{Wu}.
Given a \BF\ $(\LieG,\LieG_-,\LieG_+)$ as before, we may construct
another factorization by setting:
\BEQ
\LietilG=\LieG\times\LieG,\;\LietilG_-=\LieG_-\times\LieG_+,\;\LietilG_+=
\{(g,g)\in\LietilG\}\cong\LieG.
\EEQ
The splitting of an element $(g_1,g_2)\in\LietilG$ is given by the \BF:
If $g_1g_2\inv=g_-g_+\inv$, $\hatg=g_-\inv g_1=g_+\inv g_2$, then
\BEQ
(g_1,g_2)=(g_-,g_+)(\hatg,\hatg).
\EEQ
On the Lie algebra level this reads:
\BEQ
\Lietilg=\Lieg\oplus\Lieg,\;\Lietilg_-=\Lieg_-\oplus\Lieg_+,\;
\Lietilg_+=\diag(\Lietilg).
\EEQ
With the Killing form
$B_\tilG((x_1,x_2),(y_1,y_2))=B_G(x_1,y_1)-B_G(x_2,y_2)$ this becomes a
Manin triple.

If we restrict the $p_j$ to $\Lieg_+\oplus\Lieg_+$ then the orbit and
all flows are restricted to this subset, which is isomorphic to
$\Lieg_+$, and the
equations~\bref{UDEF} and~\bref{aux} are the same as for the \BF\
$(\LieG,\LieG_-,\LieG_+)$.
This is the reason why the flows of the mKdV systems commute with the
flows of the Leznov-Saveliev ones as mentioned in~\cite{DS}.

However, to recognize the higher symmetry of these systems, connected
with their Lorentz invariance, it is more convenient to use \Drinfeld's
double construction:\\
Given a Lie group $\LieG$ with Lie algebra $\Lieg$ being a Kac-Moody
algebra, $\Lieh$ a Cartan subalgebra and $\Lien_\pm$ the standard
nilpotent subalgebras (see below), we denote by $\LieN_\pm$
and $\LieH$ the subgroups of $\LieG$ belonging to them. We set:
\BEQ
\DG=\LieG\times\LieG,
\EEQ
\BEQ
\DG_+=\{(g,g)\in\DG\}\cong\LieG,
\EEQ
\BEQ
\DG_-=\{(n_-h,n_+h\inv)\in\DG\mid n_\pm\in\LieN_\pm,h\in\LieH\}.
\EEQ
The factorization of an element $\tilg=(g_1,g_2)\in\DG$
is achieved by using the Gauss decomposition $g_1g_2\inv=n_-h^2n_+\inv$:
If $\hatg=h\inv n_-\inv g_1=hn_+\inv g_2$, then
\BEQ
\tilg=(g_1,g_2)=(n_-h,n_+h\inv)(\hatg,\hatg).
\EEQ
On the Lie algebraic level this is the classical double $\Dg$.
Given an element $(x,y)\in\Dg$, we may write
\BEQ \label{Dgsplit}
(x,y)=(x_-+\halb x_\Lieh,x_+-\halb x_\Lieh)+(\hatx,\hatx),
\EEQ
with $x_--x_++x_\Lieh$ being the linear decomposition of $x-y$ into an
$\Lien_-$, $\Lien_+$ and $\Lieh$ part,
and $\hatx=x-x_--\halb x_\Lieh=y-x_++\halb x_\Lieh$.

Here, if $p_0$ is contained in one of the Lie subalgebras $\Lien_\pm$
(identify $\Lietilg_+$ with $\Lieg$),
we have no injectivity of $\Op$ (see also~\cite{Wu}). This can be seen
directly by looking at the expressions for
$U(\cdot)=\omega(r(p_0))(\tilg(\cdot))$:
Let the splitting of $\tilg\in\LietilG$ be as above and abbreviate $n_-h$,
$n_+h\inv$ by $g_-$, $g_+$ respectively, then
\BEQ
\omega(r(p_0))(\tilg(\unt))=q_+(\Ad(g_-\inv(\unt)) p_0)+
q_-(\Ad(g_+\inv(\unt)) p_0),
\EEQ
and the auxiliary equation splits into:
\BEQ \label{aux1}
g_-\inv(\unt)\der{g_-(\unt)}{x}
=-q_-(\Ad(g_-\inv(\unt))p_0)+q_-(\Ad(g_+\inv(\unt))p_0)
\EEQ
and
\BEQ \label{aux2}
g_+\inv(\unt)\der{g_+(\unt)}{x}=-q_+(\Ad(g_+\inv(\unt))
p_0)+q_+(\Ad(g_-\inv(\unt))p_0).
\EEQ
$q_+$ and $q_-$ being the projection given by~\bref{Dgsplit}, i.e.\
$q_\pm(x)=x_\pm\mp\halb x_\Lieh$.

For example, if $p_0\in\Lien_+$, then $U(\unt)$ does not depend on $n_+(0)$.
This failure of $\Op$ to be injective for these $p_0$ is not very
surprising, as the choice $p_0\in\Lien_\pm$ corresponds to taking starting
values on one chiral branch of the 1+1 dimensional light cone. Therefore it
can be interpreted as an expression of the non-uniqueness of the Cauchy
problem.

We will not be concerned with the question of injectivity of $\Op$ for general
$p_0$ and $G$ as before the loop group over a semisimple finite dimensional
group $\LieKC$, but let us briefly indicate the heuristical arguments leading
to the conclusion that for generic $p_0$ having non-zero parts in $\Lien_+$
and $\Lien_-$, $\Op$ is, at least
formally, injective:\\
In the following we always choose a marking of the Lie algebra
$\LiekC$ of $\LieKC$,
i.e.\ besides of the Cartan subalgebra $\LiehC$ we also choose a basis
$\Delta$ of simple roots, basis vectors $e_\alpha$ in every root space,
constituting a Chevalley basis and
nilpotent subalgebras $\Lien_\pm$, such that the positive and negative roots
are
contained in $\Lien_+$, $\Lien_-$, respectively.
We also define the Borel subalgebras $\Lieb_\pm=\Lien_\pm\oplus\LiehC$.

Using the grading of $\Lieg$ given by
$\deg(\lambda^ne_\alpha)=n+\height(\alpha)$, one sees that
the orbit is restricted to the set $\Liem_0$
of elements in $\Lieg\cong\Lietilg_+$ of which the degree does not exceed
the maximum or minimum degree of $p_0$. Therefore it will be finite
dimensional, if $p_0$ contains only finite powers in $\lambda$.
The function $U(\cdot)$ generically determines the
projection of $\log g_\pm$ to $\Liem_0$, as was the case for the NLS.

Furthermore, because the second terms on the r.h.s.\ of the
equations~\bref{aux1} and~\bref{aux2} vanish,
if we project on $\Lieg\ominus\Liem_0$, we get, also by similar
arguments as for the NLS case, recursive equations for the higher
powers of $\lambda\inv$ and $\lambda$ in $g_-$ and $g_+$, respectively.
These equations are formally solvable but in general very complicated, as they
always contain infinitely many commutators with the degree $0$ part of
$\log g_\pm$.

However, one has to be careful, whether the ansatz, being implicit in the
expression $\log g_\pm$ is justified. The requirement that the
exponential map is bijective on the small group $\LieG$ is hard to
achieve in practice. It is not true even for the modified loop group
definition given above.
The injectivity of $\Op$ is thus conserved only in an algebraic sense,
obtained by considering the universal enveloping algebras rather than
the Lie groups.

The class of algebraic geometric (finite gap)
solutions $[g_-(\unt)]$ of an integrable system is
characterized by the condition that the tangent space of the flow at the
starting point $[g_-]$ is finite dimensional, i.e.~there exist only
finitely many linear independent $p_j$ such that all their flows are
nontrivial.
In the case of the NLS systems it suffices~\cite[section~4]{HSS92}
to show that there exists one polynomial regular
$\tilp\in\Lieh_+$ such that its flow is trivial. Polynomial regularity
for an element of $L^{a+}\LiekC$ shall mean that it can be written as
a polynomial in the loop parameter $\lambda$ with leading coefficient
being a regular element of the Lie algebra $\LiekC$.
Further theorems concerning the existence of periodic algebraic
geometric solutions are given in~\cite{HSS92}. For a general account on
algebraic geometric solutions see the article of Dubrovin, Krichever and
Novikov~\cite{Novikov}.

\section{Group theoretic symmetry}
The above definition~\bref{flows} of the flows makes the following definition
of a symmetry quite natural:
\begin{Definition}
A group theoretic symmetry for an integrable system of the kind
described above is a differentiable automorphism $\phi$ of the group
$\LieG$, which
\begin{itemize}
\item[a)] respects the Birkhoff factorization, i.e.~is also an
automorphism of the subgroups $\LieG_\pm$,
\item[b)] its derivative $\diff\phi$ leaves the Cartan
subalgebra $\Lieh$ and the generators $p_0,p_1\in\Lieh$ of space and
time translation invariant.
\end{itemize}
\end{Definition}
In the case of an additional reduction, as for the KdV or mKdV systems,
this should be supplemented by a consistency condition on $\phi$.

Invariance of $\Lieh$ under $\phi$ is necessary to guarantee that $\phi$
reduces to a map on $\LieH_-\setminus\LieG_-$, which is the set of
solutions of the system.
The restriction (b) on $\phi$ can, and has to be loosened, in order to
include space-time symmetries, i.e. symmetries which do not let invariant
the flow parameters, for example the Poincar\'e symmetry, encountered in
connection with the Leznov-Saveliev systems.
We will formulate this generalization at the end of this section.

For a group theoretic symmetry $\phi$ we obviously have
\BEQ
\tilg(x,t)=\exp(-xp_0-tp_1)\tilg_-=\tilg_-(x,t)\tilg_+(x,t),
\EEQ
with $\tilg=\phi(g)$, $\tilg_\pm=\phi(g_\pm)$.
Therefore $\phi$ maps a solution of the system generated by the abstract data
$(\LieG,\LieG_-,\LieG_+)$ with starting point $g_-$ to a solution of the
same system with starting point $\tilg_-$.

As the notions of a periodic and of an algebraic geometric solution can
be expressed in a purely group theoretic way, it is easy to see that
the classes of periodic and of algebraic geometric solutions are mapped to
themselves. Every flow which acted trivial on the starting point $g_-$
is mapped to a flow which acts trivial on $\tilg_-$.
This kind of symmetry will be used to model B\"acklund transformations
for the generalized KdV equations.

If one has a Miura transformation mapping the solutions of an
integrable system (1) to another integrable system (2), being invertible
for a sufficiently large class of solutions of (2), and a symmetry
of the second one, then it is possible to ``pull back'' this symmetry by the
Miura transformation to a symmetry of the first one.
This transformation we call a B\"acklund transformation.

Clearly, if the Miura transformation would constitute an isomorphism of
Birkhoff factorizations, the B\"acklund transformations would not be of
any interest, because then the symmetries of the systems could be
easily translated into each other by purely group theoretic means.
But for the generalized KdV equations and their counterpart,
the generalized mKdV equations of \Drinfeld\ and Sokolov, this is not the case.
There the Miura transformation is the projection of one coset space into
another~\cite{DS,HSS92}, which can have a very complicated
form if being expressed in coordinates of the Lie-Algebra $\Lieg_+$.
For the groups $\LieG=\calL\LieSL\nC$, calculating the inverse of the
Miura transformation in general amounts to solving a system of
$n-1$ differential equations of order $n-1$ and degree $n$.

In the following we will elaborate on this standard example of a
B\"acklund transformation.

The following reduction of the generalized NLS systems results in a
generalization
(not the one of \Drinfeld\ and Sokolov) of the modified KdV equations:\\
Let $\Liehatg\subset\Lieg$ be the covering Lie algebra of $\LiekC$
generated by the involution (notation as before):
\BEQ
\sigma: \left\{\begin{array}{l}
          \LiehC\ni x \mapsto -x,\\
          e_\alpha \mapsto e_{-\alpha}\;\forall\alpha\in\Delta
        \end{array}\right.
\EEQ
So $\Liehatg$ is generated by $\lambda^{2n+1}\LiehC$, and the elements
$\lambda^{2n+1}(e_\alpha-e_{-\alpha})$,
$\lambda^{2n}(e_\alpha+e_{-\alpha})$, $\alpha\in\Delta$, $n\in\Bnn$.

The parameter-dependent modified KdV equation for $\LieKC=\LieSL(2,\Bcc)$,
\BEQ
\psi_t=6(\psi^2+c)\psi_x-\psi_{xxx},
\EEQ
$c$ being an arbitrary complex parameter,
which was used by Miura~\cite{Miura68}
cannot be obtained by a group theoretic reduction of $\Lieg$.
It rather connects with a dynamical restriction of the flows to an affine
subspace
\BEQ \label{affres}
\Liem=\tilc\sigma_2+\Liehatg,\;\tilc=\sqrt{-c}
\EEQ
of \Lieg.
One may say that, by an additional symmetry of the third order flow of
the NLS hierarchy, which yields the mKdV equation, a solution stays in
$\Liem$, when its starting point $g_-(x,0)$ was inside $\Liem$ for all $x$.
Neither the existence of such a starting point, nor the above fact can
be easily proved by group theoretic arguments.

However, we know that there exists a Miura transformation for every
choice of the parameter $c$
connecting the corresponding mKdV with the KdV equation:
\BEQ
u=\psi_x+\psi^2+c.
\EEQ
We will now show, that there exists an analogous restriction of the
generalized mKdV equations of \Drinfeld\ and Sokolov
for higher groups $\LieKC=\LieSL\nC$ together with a group
theoretic symmetry generalizing the construction for $\LieSL(2,\Bcc)$.
Let us therefore recall the procedure of~\cite{DS}, which gives the
generalized Miura transformations.

We restrict the group $\LieG$ to the subgroup $\LiechG$ of loops
in $L^a\LieK_\Bcc$ which
factorize through the map $\lambda\mapsto\lambda^n$ of $\LieG$ onto itself,
where $n-1$ is the
height of the unique highest root $\delta$, w.r.t.\ the chosen basis
of simple roots.
Let $\Liechg\subset\Lieg$ be the Lie algebra of $\LiechG$, set
$z=\lambda^n$ and choose the splitting
\BEQ
\Liechg_-=(\Lieg_-\oplus\Lien_-)\cap\Liechg,
\EEQ
\BEQ
\Liechg_+=(z\Lieg_+\oplus\Lieb_+)\cap\Liechg
\EEQ
of $\Liechg$ and the momentum generator
\BEQ
q_0=\sum_{\alpha\in\Delta}e_\alpha+ze_{-\delta}\in\Liechg_+,
\EEQ
i.e.\ the subalgebra $\Liechg_-$ is enlarged, compared to the Birkhoff
factorization of the NLS.
By the procedure described in the section before, one gets
an injective map $\Oq:\LiechG_-(q_o)\setminus\LiechG_-\rightarrow
q_0+\calK_0(\Brr,\Lieh_\Bcc)$.

Let $\Lieu\oplus[q_0,\Lien_-]$ be a vector space decomposition of the Borel
subalgebra $\Lieb_-$.
In the case of $\LieK=\Liesl\nC$, $\Lieu$ is the set of $\Liesl\nC$ matrices
having non-zero entries only on the lowest row.

We now view $\calK_0(\Brr,\Lieh_\Bcc)$ as a subspace of
$\calK_0(\Brr,\Lieb_-)$.
The set of functions $\calK_0(\Brr,\LieN_-)$, $\LieN_-$ the group corresponding
to $\Lien_-$, acts as a right transformation
group on the latter by
\BEQ
v'(x)=\Ad(n_-\inv(x))v(x)+n_-\inv(x)\der{n_-(x)}{x},\;
n_-(\cdot)\in\calK_0(\Brr,\LieN_-),
\EEQ
$v(\cdot)\in\calK_0(\Brr,\Lieb_-)$.
In \cite{HSS92} it was shown that by this action,
$\Oq$ projects down to an injective map
\BEQ
\Pq:\chO_{q_0}=\LiechG_-(q_0)\setminus\LiechG_-/N_-\rightarrow
q_0+\calK(\Brr,\Lieu).
\EEQ
This map plays the r\^ole of the inverse scattering problem of the KdV
equation. The projection
\BEQ \label{proj}
\LiechG_-(q_0)\setminus\LiechG_-\rightarrow\chO_{q_0}
\EEQ
is the generalized Miura transformation of \Drinfeld\ and Sokolov.

The triple $(\Liechg,\Liechg_-,\Liechg_+)$ is a reduction (in the sense
of~\cite{HSS92})
of the Manin triple $(L^a\LiekC,{}^-L^a\LiekC,{}^+L^a\LiekC)$, where
${}^-L^a\LiekC$ is generated by the Cartan subalgebra of $\LiekC$ and
all elements $\lambda^ke_\alpha$ with $\height(\alpha)+k>0$, and
${}^-L^a\LiekC$ is generated by all such elements with $\height(\alpha)+k<0$.
The latter may be transformed first by an automorphism
\BEA
\lambda^ke_\alpha & \mapsto & \lambda^{\height(\alpha)+k}e_\alpha,\nonumber \\
\Lieh_\Bcc\ni x & \mapsto & x,
\EEA
then by the inner automorphism of $\Liek_\Bcc$ conjugating $q_0$
into the Cartan subalgebra $\LiehC$.
This is a transformation of the above Manin triple to the
triple of the NLS equation, mapping the momentum generator of the mKdV system
$q_0$ to the momentum generator $p_0$ of a system of NLS type.
Thus the generalized mKdV systems of \Drinfeld\ and Sokolov are also
embedded in the NLS hierarchy.

The above construction relating solutions of the mKdV and KdV systems
may be varied for $\LieK_\Bcc=\LieSL\nC$ in the following way:

Instead of identifying the spectral parameters of the KdV and mKdV model as
above, we set
\BEQ
z=\tilz-c^n,\;c\in\Bcc,
\EEQ
where $z$ and $\tilz$ are the spectral parameters of the KdV and mKdV
equations, respectively.

Let now $a_j$, $j=0,\ldots,n-1$ be the roots of the equation
$\lambda^n=c^n$,
\BEQ
a_j=ce^{2\pi i(n-j)\over n}.
\EEQ
This gives
\BEQ
\prod_{j=0}^{n-1}(\lambda-a_j)=\lambda^n-c^n.
\EEQ
We choose the standard basis of simple roots
$\{\alpha_1,\ldots,\alpha_{n-1}\}$ of $\Liesl\nC$,
$\alpha_j=e_{jj}-e_{j+1,j+1}$, $e_{jk}$ being the matrix units
$(e_{jk})_{lm}=\delta_{jl}\delta_{km}$.

Setting $\tilz=\lambda^n$ and performing the automorphism
\BEA
e_{\alpha_j} & \mapsto & (\lambda-a_j)e_{\alpha_j}, \nonumber\\
e_{-\alpha_j} & \mapsto & (\lambda-a_j)\inv e_{-\alpha_j}, \nonumber\\
\Lieh_\Bcc\ni x & \mapsto & x,
\EEA
out of $q_0$ we obtain the element
\BEQ
\tilq_0=\sum_j(\lambda-a_j)e_{\alpha_j}+(\lambda-a_0)e_{-\delta}\DEF
\lambda A+\tilC.
\EEQ
The diagonalization matrix of $A$ is
$(Q)_{jk}=\sqrt{1\over n}\param^{jk}$, $\param=e^{2\pi i\over n}$,
all matrix indices now running from $0$ to $n-1$.
Let $B$ be a matrix which has entries only on one diagonal, i.e.
\BEQ
(B)_{jk}=\left\{\begin{array}{ll} b_j & \mbox{if $k=j+\epsilon$
                                              or $k+n=j+\epsilon$,}\\
0 & \mbox{otherwise,}
\end{array}\right.
\EEQ
for some integer constant $0\leq\epsilon\leq n-1$.
The conjugation with $Q$ maps $B$ to
\BEQ
(Q\inv BQ)_{jk}=\param^{k\epsilon}\sum_{m=0}^{n-1}b_m\param^{m(k-j)},
\EEQ
The matrix $\tilC$ therefore changes to
\BEQ
C=-c\left(\sum_{j=1}^{n-1}\param^{j-1}e_{\alpha_j}+\param^{n-1}
e_{-\delta}\right),
\EEQ
i.e.\ the matrix $\tilq_0$ takes the form
\BEQ
p_0+C=\sum_{j=0}^{n-1}\param^je_{jj}+C,
\EEQ
where $p_0$ is the $x$-flow generator of the NLS systems.

The diagonal matrices, containing the fields, are transformed into
off-diagonal matrices. This means that the above chain of
transformations leads again to an element of $\Op$ but now not
contained in the section $\Op\cap\Liehatg$.
It also maps $\Liechg_-$ and $\Liechg_+$ to the \BF\ $\Lieg_-$ and
$\Lieg_+$ of the NLS models.

In analogy to the case $\LieKC=\LieSL(2,\Bcc)$ we now restrict the flows
of the generalized NLS systems to the image of the generalized KdV systems
under the above constructed Miura transformation depending on the parameter
$c$. The matrix $C$ is therefore a generalization of the matrix
$\tilc\sigma_2$ in \bref{affres}.

As mentioned before, a B\"acklund transformation is modeled out of a
Miura transformation between two integrable systems and a symmetry.
We are therefore looking for an automorphism leaving $p_0+C$ and the
image of $\LiehC$, in which the fields are contained, invariant.
In addition it has to respect the \BF\ of the generalized NLS equations.

An obvious possibility is the following linear transformation of Chevalley
generators:
\BEA
e_{\alpha_j} & \mapsto & \param e_{\alpha_{j+1}},\;
j=1,\ldots,n-2,
\nonumber \\
e_{-\alpha_j} & \mapsto & e_{-\alpha_{j+1}},\;
j=1,\ldots,n-2,
\nonumber \\
e_{\alpha_{n-1}} & \mapsto & \param e_{\alpha_{-\delta}}, \nonumber\\
e_{-\alpha_{n-1}} & \mapsto & e_{\alpha_\delta} \label{varphi}
\EEA
and the resulting transformation of the Cartan subalgebra.
We will not write down the general transformation of the fields explicitly.

In the case $n=2$ this all reduces to the well known procedure
for the KdV system.
It also reproduces the parameter-dependent Miura transformation, which
can be calculated explicitly.

In the case $n=3$ we obtain the following explicit formulas
($\param=e^{2\pi i\over 3}$):
\BEQ
q_0=\left(\begin{array}{ccc} 0 & 1 & 0 \\
0 & 0 & 1 \\
\tilz & 0 & 0 \end{array}\right),
\EEQ
\BEQ
\tilq_0=\left(\begin{array}{ccc} 0 & \lambda-c\param^2 & 0 \\
0 & 0 & \lambda-c\param \\
\lambda-c & 0 & 0
\end{array}\right),
\EEQ
\BEQ
p_0+C=\lambda\left(\begin{array}{ccc} 1 & 0 & 0 \\
0 & \param & 0 \\
0 & 0 & \param^2\end{array}\right)-\left(\begin{array}{ccc} 0 & c & 0 \\
0 & 0 & c\param \\ c\param^2 & 0 & 0
\end{array}\right).
\EEQ
The Miura transformation becomes more complicated, and is not explicitly
solvable, since it is no more of Riccati type. We give it for $c=0$. If
\BEQ
q_0+u=\left(\begin{array}{ccc} 0 & 1 & 0 \\
0 & 0 & 1 \\
u_1+z & u_2 & 0 \end{array}\right)
\EEQ
is an element of $\chO_{q_0}$ and
\BEQ
U=\left(\begin{array}{ccc} \psi & 1 & 0 \\
0 & \phi-\psi & 1 \\
z & 0 & -\phi \end{array}\right)
\EEQ
an element of $\LiechG(q_0)\setminus\LiechG$ which is mapped by the
projection~\bref{proj} to $q_0+u$, then
the Miura transformation for the fields is
\BEQ
u_2=(\phi+\psi)_x+\psi^2-\psi\phi+\phi^2
\EEQ
and
\BEQ
u_1=-u_2\psi+r_x+(\psi+\phi)r+\halb\psi\phi(\phi-\psi),
\EEQ
where $r=\psi_x-\halb\psi(\phi-2\psi)$.
The symmetry transformation of the fields is
\BEA
\psi & \mapsto & \widetilde{\psi}=s^2\phi, \nonumber \\
\phi & \mapsto & \widetilde{\phi}=-s\inv\psi-s\phi.
\EEA
It is not clear, that the resulting
B\"acklund transformation maps algebraic geometric into algebraic geometric
solutions. Periodicity is not conserved in general. This has to be proven
in every case separately by looking at the resulting differential equations.
For the KdV equation with periodic boundary conditions, the B\"acklund
transformation just amounts to adding one more gap to the spectrum of Hill's
equation, or equivalently one more singularity to the Riemannian surface of
an algebraic geometric solution~\cite{HSS92}. A B\"acklund
transformation for the group theoretic systems should therefore conserve
the quality of algebraic geometricity.

The class of time-flow generators $q_1$, which are
mapped by the above parameter-dependent transformations to elements
commuting with $p_0$ is restricted.
An example for such a generator is
\BEQ
\tilq=(z+c^n)q_0.
\EEQ
It is mapped to $\lambda^np_0$. This means that we must investigate the
mKdV equation for $q_0$, $q_1=zq_0$ with the transformed space-variable
$\hatx=x+c^nt$:
\BEQ
\exp-(xq_0+t\tilq)=\exp-((x+c^nt)q_0+tq_1).
\EEQ
Thus a stationary solution for $c=0$ will be a wave propagating
with velocity $-c^n$ for general $c$.

The symmetry automorphism~\bref{varphi} together with the generalized
parameter-dependent Miura transformation yields a B\"acklund
transformation for the group theoretically generalized KdV systems in the
case of $\LiekC=\Liesl\nC$. For other simple Lie algebras
a similar construction should be possible by taking matrix
representations which are reductions of the defining representation of an
$\Liesl\nC$.

\bigskip To show that the above definition of a symmetry is also of
independent value, we
provide at last the promised formulation of space-time symmetries of
group theoretically describable systems.\\
If one chooses two generators $p_0$, $p_1$ for space and time flow,
respectively, then in order to maintain the ZS type equation~\bref{ZS}
for $V_0$ and $V_1$, it is enough to leave the scalar product
$\lpair\unt,\unp\rpair=xp_0+tp_1+\ldots$ in~\bref{flows} invariant.
When transforming $p_0$ and $p_1$ nontrivially, we must therefore also
change the parametrization of space-time. The picture is similar to the
``passive'' one in special relativity.

An example for such a symmetry is provided by the implementation of the
Lorentz group action in the group theoretic scheme for the Leznov-Saveliev
systems. As we are working in light-cone coordinates, we write $\xi=t_0$
and $\eta=t_1$.
There we have the following automorphisms:\\
The Cartan involution $T$ of the loop group,
\BEQ
e_\alpha\rightarrow-e_{-\alpha},\;\lambda\rightarrow\lambda\inv,\;
\Lieh\rightarrow-\Lieh
\EEQ
maps $p_0$ for the LS systems to $-p_1$ and vice versa. $T$ therefore
amounts to the time reversal $\xi\leftrightarrow-\eta$.

The additional multiplication by $-1$ gives a space reversal
$\xi\leftrightarrow\eta$.
The boosts $\xi\mapsto\kappa\xi$, $\eta\mapsto\kappa\inv\eta$
are implemented by the automorphism
\BEA
e_\alpha & \mapsto & \kappa e_\alpha,\;\alpha\in\Delta^+, \nonumber\\
e_{-\alpha} & \mapsto & \kappa\inv e_{-\alpha},\;\alpha\in\Delta^+,
\nonumber\\
\lambda & \mapsto & \kappa^n\lambda,
\EEA
which is the identity on $\LiehC$.

\separate The author thanks M. Schmidt and R. Schrader for helpful
discussions on the subject. This work was supported by Deutsche
Forschungsgemeinschaft as part of Sonderforschungsbereich 288
``Differentialgeometrie und Quantenphysik''.

\end{document}